\documentclass[journal=jpccck,manuscript=article]{achemso}
\usepackage[version=3]{mhchem} 
\usepackage{natmove}
\usepackage{multirow}
\usepackage{array}
\usepackage{rotating}
\usepackage{longtable}
\usepackage{pdfpages}
\author{Caterina Cocchi}
\email{caterina.cocchi@unimore.it}
\affiliation{Centro S3, CNR-Istituto Nanoscienze, I-41125 Modena, Italy}
\alsoaffiliation{Dipartimento di Fisica, Universit\`a di Modena e Reggio Emilia, I-41125 Modena, Italy}
\author{Deborah Prezzi}
\affiliation{Centro S3, CNR-Istituto Nanoscienze, I-41125 Modena, Italy}
\email{deborah.prezzi@unimore.it}
\author{Alice Ruini}
\affiliation{Centro S3, CNR-Istituto Nanoscienze, I-41125 Modena, Italy}
\alsoaffiliation{Dipartimento di Fisica, Universit\`a di Modena e Reggio Emilia, I-41125 Modena, Italy}
\author{Marilia J. Caldas}
\affiliation{Instituto de F{\'\i}sica, Universidade de S\~ao Paulo, 05508-900 S\~ao Paulo, SP, Brazil}
\author{Elisa Molinari}
\affiliation{Centro S3, CNR-Istituto Nanoscienze, I-41125 Modena, Italy}
\alsoaffiliation{Dipartimento di Fisica, Universit\`a di Modena e Reggio Emilia, I-41125 Modena, Italy}
\title{Electronics and Optics of Graphene Nanoflakes: Edge Functionalization and Structural Distortions}
\begin{document}
\begin{abstract}
The effects of edge covalent functionalization on the structural, electronic 
and optical properties of elongated armchair graphene nanoflakes (AGNFs) are 
analyzed in detail for a wide range of terminations, within the framework of 
Hartree-Fock-based semi-empirical methods.
The chemical features of the functional groups, their distribution and the 
resulting system symmetry are identified as the key factors that determine the 
modification of structural and optoelectronic features.
While the electronic gap is always reduced in presence of substituents, 
functionalization-induced distortions contribute to the observed lowering by 
about 35-55$\%$. This effect is paired with a red shift of the first optical 
peak, corresponding to about 75$\%$ of the total optical gap reduction.
Further, the functionalization pattern and the specific features of the edge-%
substituent bond are found to influence the strength and the character of the 
low energy  excitations. 
All these effects are discussed for flakes of different width, representing the 
three families of AGNFs.
\end{abstract}

\textbf{Keywords}:  ZINDO, AM1, substitution, UV-vis spectrum, graphene nanoribbons, configuration interaction

\newpage
The electronic \cite{cast+09rmp,naka+96prb,son+06prl,baro+06nl}, optical 
\cite{prez+08prb,yang+07nl,yang+08prl} and magnetic properties 
\cite{yazy10rpp,pisa+07prb,son+06nat} of graphene nanoribbons are closely related 
to their edge morphology and indeed several theoretical proposals have been 
advanced to tune the properties of these systems through edge modifications.
For instance, the termination of edge C atoms with different groups 
\cite{hod+07nl,gunl+07apl,kan+08jacs,wu+10jpcc} and the introduction of magnetic 
impurities \cite{rigo+09prb,long+10prb,wang+10prb,cocc+10jcp} at the edges of 
zigzag graphene nanoribbons represent promising routes to fully exploit their 
potential for spintronics applications.
In armchair graphene nanoribbons (AGNRs), on top of the intrinsic tunability 
given by the well-established family behavior \cite{naka+96prb,son+06prl}, edge 
modifications can further engineer the energy gap and work function 
\cite{cerv+07prb,ren+07condmat,lu+09apl,wang-li11jap,cocc+11jpcc}, as well as 
their optical response \cite{zhu-su10jpcc,prez+11prb,cocc+11jpcl}.

An additional aspect of edge modifications is related to the appearance of structural 
distortions: ripples and twists can arise, either localized at the edge region 
\cite{wagn+11prb,rose+11prb} or extended to the whole structure, with effects on 
a larger scale \cite{bets-yako09nr,shen+10nano,lu-hung10prb}. 
Indeed, the study of structural modifications has drawn itself great attention, 
since they can lead to a futher tuning of the graphene nanostructure properties 
\cite{peng-vela11apl,sadr+11apl}. 
We here focus on unraveling the effects of functionalization-induced structural, 
electronic and optical modifications, considering not only the functional group 
itself but also the functionalization pattern. 
Such analysis can be particularly relevant to design new optoelectronic 
functionalities in graphene nanostructures, above all in light of the recent 
improvements in their controlled synthesis at the nanoscale
\cite{wu+07cr,kosy+09nat,jiao+09nat,cai+10nat,jia+10ns,palm-samo11natc,blan+12nano}.

We carry out our analysis on short armchair graphene ribbons functionalized with
different edge-substituents, ranging from single atoms to small molecular moieties.
Family-dependence is also taken into account.
We recognize that several factors --including the system symmetry, the chemical
features of the functional groups and their distribution-- are responsible for
the appearance of different structural and optoelectronic modifications.            
In general, the steric hindrance of substituents is found to induce local 
distortions at the edges, but global effects involving the flake backbone can 
also arise in specific cases.
These structural modifications are found to reduce the electronic gap up to 300 meV,  
corresponding to 55$\%$ of the total gap decrease. This effect is associated to 
a similar red-shift of the first optical peak.
On the other hand the chemical features of the substituents and their arrangement
at the edges crucially influence the distribution of the frontier orbitals. 
Local variations arise as a consequence of the modified molecular symmetry upon edge 
functionalization, while global localization effects are noticed when a large longitudinal 
dipole component is introduced by polar substituents. 
Moreover, such charge redistribution results in the appearance of an additional peak in 
the low energy region of the spectra, related to the optical activation of an otherwise 
dark excitation.

%
\section{Methods and System Description}
The results presented in this paper are obtained within the framework of 
Hartree-Fock based semi-empirical methods \bibnote{AM1 and ZINDO/S calculations 
were performed using VAMP package included in Accelrys Materials Studio software, 
version 5.0  (\url{http://accelrys.com/products/materials-studio}).}, which are 
well tested and reliable for the evaluation of the electronic and optical 
properties of C-conjugated low-dimensional systems \cite{wetm+00cpl,cald+01apl,%
davi-cald02jcc,kuba+02nat}.
The AM1 model \cite{dewa+85jacs} is adopted for structural optimization 
(0.4 $\text{kcal} \cdot \text{mol}^{-1}$/\AA{} threshold for the forces) and for 
the study of the electronic properties, including the calculation of electric 
dipole moments.
The electronic gap is computed as the difference between \textit{vertical} 
ionization potential (IP) and electron affinity (EA), which are in turn obtained 
from the total energy of the neutral and ($\pm$1) charged states, i.e. IP = E(+1) 
- E(0) and EA = E(0) - E(-1).
The optical spectra are evaluated by means of the ZINDO/S model \cite{ridl-zern73tca}, 
with single excitation Configuration Interaction (CIS).
Our convergence tests over the number of occupied and virtual molecular orbitals 
(MOs) indicated that a CI energy window of at least 4.5 eV below the HOMO and 
3.5 eV above the LUMO is required for a reliable characterization of the low 
energy optical excitations.

\begin{figure}%
\centering
\includegraphics[width=.45\textwidth]{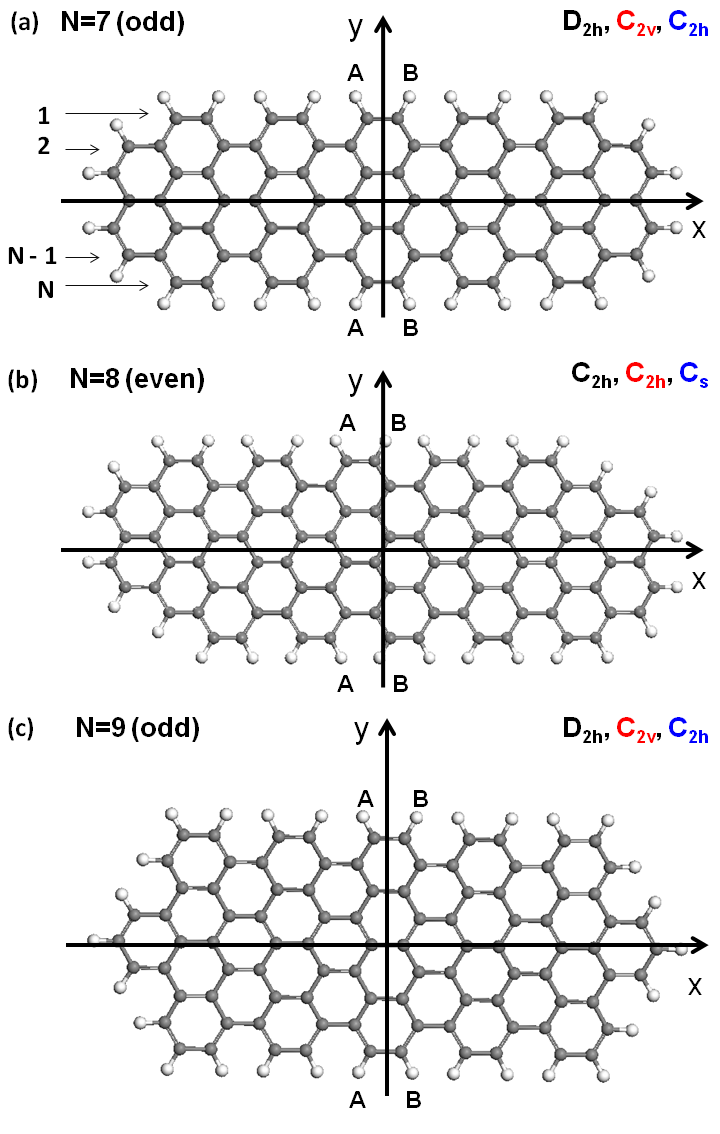}%
\caption{H-terminated armchair graphene nanoflakes (AGNFs) of 
fixed length ($x$ axis) and variable width ($y$ axis), chosen as representative 
of the three different families of AGNFs, i.e. $N$=$3p$, $3p+1$ and $3p+2$.
Edge functionalization is performed by replacing each second H atom along the 
length with a foreign substituent. Two functionalization schemes can be identified: 
the A-A scheme, where the same zigzag line is functionalized on both edges, and 
the A-B scheme, where functionalization involves alternating zigzag lines. 
Indicated are the point symmetry groups of each structure, clean (black), and 
functionalized in the A-A (red) and A-B (blue) scheme, respectively, for ideal 
(non-distorted and symmetrically terminated) flakes.
}
\label{fig1}
\end{figure}
%

We perform our analysis on elongated graphene nanoflakes (GNFs) with armchair-%
shaped edges, which can be seen as finite portions of graphene nanoribbons. As
for infinite ribbons, we introduce a width parameter $N$, corresponding to the 
number of dimeric C lines across the $y$ axis (see \ref{fig1}a). According to 
their width parameter $N$, both ribbons \cite{naka+96prb,son+06prl} and flakes
\cite{cocc+11jpcc} present three scaling laws for the energy gap, also in presence 
of edge substituents. This allows us to classify them into three different families, 
following a \textit{modulo 3} periodic law: $N=3p+m$, with $p$ integer and $m=0,1,2$.
In the following, we will mainly focus on a $N=7$ GNF (width $\sim$ 7.3 \AA) of 
length $\sim$ 24 \AA{} ($x$ axis, see \ref{fig1}a). 
We will then extend our analysis to the family dependence by considering two 
additional structures (shown in \ref{fig1}b-c), having widths 8.6 \AA{} ($N$=8) 
and 9.8 \AA{} ($N$=9) and fixed length. 
For each structure, the lateral ends along the $x$ axis are shaped in such a way 
to minimize the influence of zigzag edges, which are known to affect the electronic 
properties of the system \cite{shem+07apl,hod+08prb}.

\begin{figure}
\centering
\includegraphics[width=.95\textwidth]{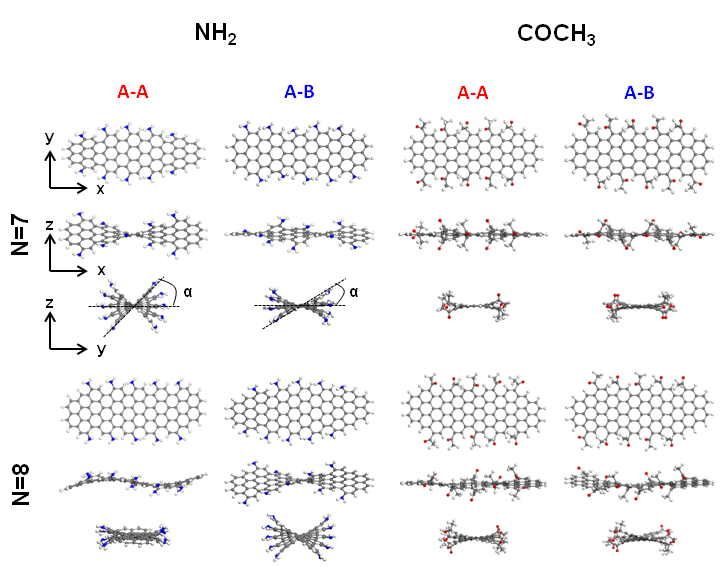}%
\caption{Structurally optimized $N$=7 and $N$=8 AGNFs, functionalizalized with 
\ce{NH2} (left) and \ce{COCH3} (right) groups, according to the A-A and A-B schemes. 
The angle $\alpha$ is introduced to quantify the distortion angle, defined as the 
maximum deviation from planarity of edge C atoms.}
\label{fig2}
\end{figure}
%

The effects of edge substituents on the electronic and optical properties of GNFs 
are investigated for a number of light-element metal-free functional groups, 
including atoms and molecular groups typically involved in electrophilic aromatic 
substitutions \cite{bade-chan89jpc,hehr+72jacs}.
We group these terminations into three main classes, with respect to the nature 
of their covalent bond with the graphene edge: the first class comprises monovalent 
halogen terminations (F, Cl and Br); the second consists of groups forming a 
direct bond at the edge, in which either N or O shares a $2p$ lone pair with 
the edge C atom (\ce{NH2}, \ce{OCH3} and \ce{OCF=CF2}); as a third class, we 
finally consider small moieties bonded to the GNF through a C-bond and including 
a C-double bond, which allows to extend the C-conjugation of the graphene backbone 
for one additional bond length beyond the edge (\ce{COCH3}, \ce{CH=CH2} and 
\ce{CH=CF2}).
The choice of these terminations is basically motivated by their experimental 
feasibility in systems similar to those investigated here \cite{zhan+05ol,%
kiku+07ol,wu+07cr,wang+09sci}.

Edge covalent functionalization is here performed by replacing each second H 
atom passivating the armchair edges in the reference system, through the regular 
alternation of substituents.
As indicated in \ref{fig1}, two functionalization schemes can be designed: in 
the A-A scheme the same C zigzag line is functionalized on opposite edges, while
in the A-B scheme alternating C zigzag lines are functionalized. In addition,
full substitution, i.e. functionalization of all C edge atoms, is considered for 
monoatomic halogen terminations only, as justified by their reduced size.
Note that, depending on the chosen scheme and on the flake width, each structure 
has different symmetry. H-terminated flakes have D$_{2h}$ (C$_{2h}$) symmetry 
for $N$ odd (even), which is modified by the presence of functional groups at the 
edges. For $N$ odd, the A-A scheme results in a C$_{2v}$ symmetry, while in the 
A-B scheme the symmetry is C$_{2h}$. On the contrary, for $N$ even the symmetry 
is preserved as C$_{2h}$ in the A-A scheme and lowered to C$_s$ in the A-B scheme. 
Even though this classification is formally correct only for ideal non-distorted
and symmetrically terminated structures, since the symmetry is always lowered 
upon structural optimization, we will keep this notation throughout the paper 
for convenience.

\section{Results and Discussion}

Edge covalent functionalization is known to be an effective strategy to tune the 
electronic gap of graphene nanostructures \cite{cerv+07prb,ren+07condmat,%
peng-vela11apl,wang-li11jap}.
As suggested already for simple aromatic compounds \cite{bock+85jmst,camp+03jpca,%
kryg-step05cr}, the presence of edge functional groups modifies the electronic 
distribution within the graphene network in the edge proximity.
As a consequence, heterospecies with different electronegativity than C and 
covalently bonded at the edges of a graphene ribbon allow to modify the ionization 
potential (IP) and the electron affinity (EA) of the system with respect to the 
reference H-passivated flake \cite{cocc+11jpcc}. 
The optical characteristics are then directly affected \cite{cocc+11jpcl}.
However, the edge-decoration of the flakes induces also structural modifications, 
which are closely related and interdependent with the electronic stabilization, 
as we will see in the following.

\ref{fig2} shows that, depending on the value of $N$ (even or odd), the type of 
substituent and the functionalization scheme (A-B or A-A), different types of 
distortions can arise, either localized at the edges or extended to the flake 
backbone. 
For instance, for \ce{COCH3} we have a typical case of structural modifications
mostly localized at the edges: there is a slight difference for different $N$ 
and scheme, however there are no global, coherent modifications. 
In this case, local steric effects among the functional groups play a primary 
role in driving distortions: \ce{COCH3} groups, which can reorient to minimize 
the repulsion, induce relatively small tilt of edge C rings to opposite directions, 
perpendicular to the flake basal plane ($z$ axis, see \ref{fig2}). 
As a second example, we highlight the case of the \ce{NH2} termination: here the 
interplay of functionalization pattern, group rigidity and steric effects can 
produce coherent distortions involving the whole flake backbone, which thus 
stabilize the structure.
This is evident for $N$=7 ($N$=8) in the A-A (A-B) scheme (\ref{fig2}), where the 
system tends to twist along its longitudinal axis (x). For $N$=8 in the A-A scheme, 
a ``rippling'' mode tends also to appear on top of the local distortions induced 
at the edges. 

To quantify these effects, we introduce an angle $\alpha$, defined as the maximum 
deviation of edge C atoms with respect to the reference planar configuration 
(H-GNF, $\alpha$=0).
The values of $\alpha$ for different functional groups are reported in \ref{table2} 
for $N$=7 GNFs (see Supporting Information for $N$=8 and $N$=9, Table S3). 
In the case of halogen termination, similar distortion angles are observed for 
both A-A and A-B schemes, while a significantly larger tilting angle is produced 
upon full functionalization.
For molecular substituents, where larger distortions are expected in view of 
their size, the tilt of edge C-rings is considerably reduced when the groups 
are relatively free to rearrange, as we described for \ce{COCH3}. This is also 
the case of methoxy (\ce{OCH3}) and \ce{OCF=CF2} groups, where the flexible 
O-linkage allows them to assume an ordered alignment (see Supporting Information, 
Figure S1). On the contrary, larger $\alpha$ angles (and global distortions) are 
noticed for rigid groups like \ce{NH2} and ethene units (\ce{CH=CF2} and 
\ce{CH=CH2}.
We remark finally that, even when we find discernible structural modifications,
they do not result in any significant local deviation from the regular graphene 
bonding and no strongly localized defects appear.
For further structural details (edge C-C and C-substituent bond lengths), see
Supporting Information, Table S1-S2.

\begin{table*}
\begin{tabular}{c||c|c|c||c|c|c||c|c|c}
\hline \hline
 & \multicolumn{3}{|c||}{\textbf{Halogen}} & \multicolumn{3}{c||}{\textbf{Direct bond}} & \multicolumn{3}{|c}{\textbf{C bond}} \\ \hline
 & F & Cl & Br & \ce{NH2} &  \ce{OCH3} & \ce{OCF=CF2} & \ce{COCH3}  & \ce{CH=CF2} & \ce{CH=CH2} \\ \hline
A-A  & 7.2  & 11.3  & 12.0    & 36.5 & 11.1   & 22.2   & 10.5 & 19.2 & 25.3 \\ \hline
A-B  & 6.6  & 10.6 & 12.5 & 22.8 & 8.7 & 15.3 & 12.5 & 26.2 & 15.5  \\ \hline
full & 17.1 & 31.3 & 29.1 & -   & -   & -    &  -   & -    & -  \\ 
\hline \hline
\end{tabular} 
\caption{
Distortion angles $\alpha$ (in degrees) for $N$=7 GNFs functionalized according 
to the indicated schemes ($1^{st}$ column). The angle $\alpha$ indicates the 
maximum deviation of edge C atoms as compared to the reference planar H-GNF, 
as shown in \ref{fig2}.
}
\label{table2}
\end{table*}

\begin{figure*}
\centering
\includegraphics[width=.95\textwidth]{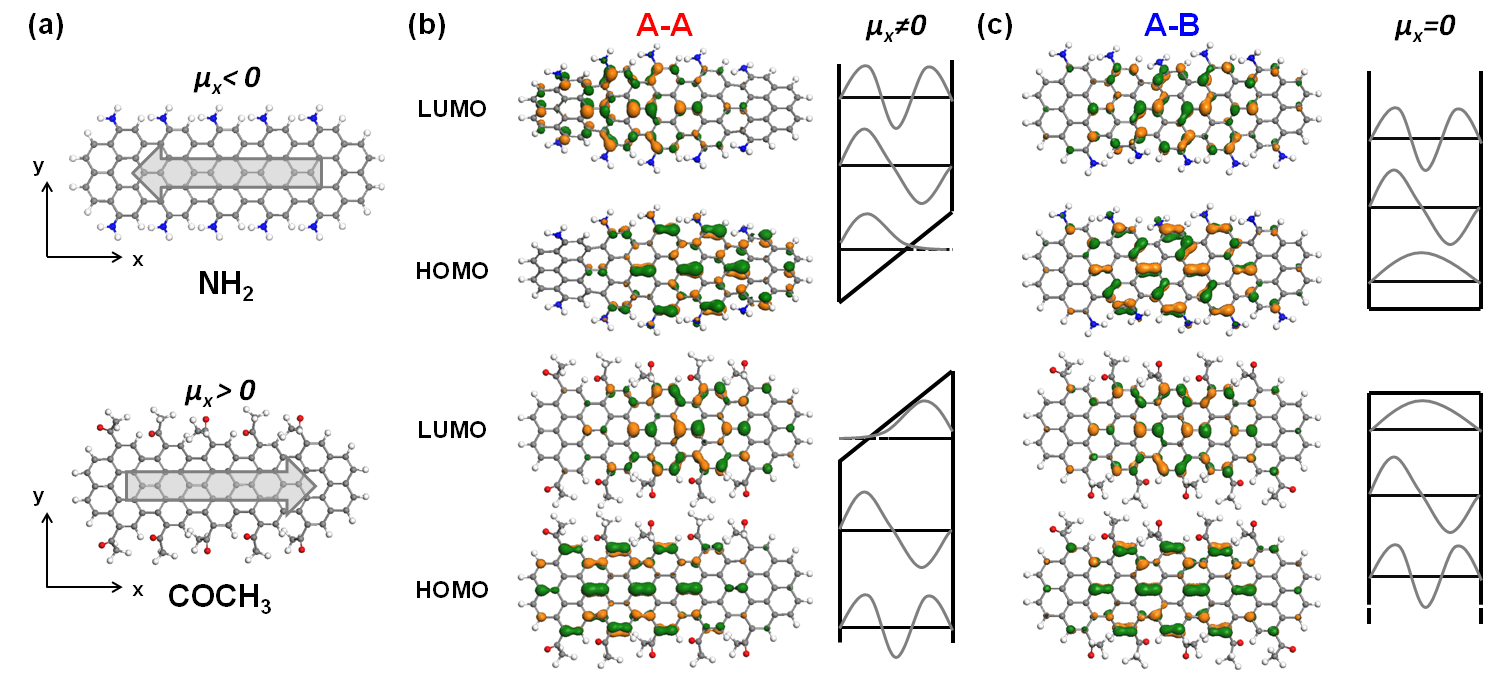}%
\caption{
(a) A-A functionalization of $N$=7 GNF by means of electron-donating (e.g \ce{NH2}) 
and -withdrawing (e.g. \ce{COCH3}) groups gives rise to longitudinal dipoles 
($\mu_x$) oriented in opposite directions. 
The frontier orbitals of \ce{NH2}- and \ce{COCH3}-substituted $N$=7 GNFs are shown 
for the (b) A-A and (c) A-B schemes. The square well potential profiles modulating 
the molecular orbital distribution are sketched in presence (b) and absence (c) of 
a longitudinal dipole along the flake.
}
\label{fig3}
\end{figure*}

We now move to the electronic properties and list in \ref{table1} the total 
dipole arising in presence of different substitution patterns.
We notice that a large dipole along the flake longitudinal axis (x) appears
for $N$ odd (here $N$=7) in the A-A functionalization scheme, while it is 
almost negligible for the other scheme. The situation is reversed for $N$ 
even (e.g. $N$=8, see Supporting Information, Table S4). 
As shown in \ref{table1} and in \ref{fig3}a, the dipole orientation and 
intensity depend on the characteristics of the functional group: electron-donating 
(e.g. \ce{NH2}) and -withdrawing groups (e.g. F and \ce{COCH3}) produce large 
longitudinal dipoles oriented in opposite directions (see also \ref{fig3}a).
The presence of a sizeable longitudinal dipole (A-A pattern for $N$=7 GNF) is 
accompanied by a spatial localization of the frontier orbitals on opposite sides 
of the flake. 
As depicted in \ref{fig3}b, for $\mu_x \ne 0$ the potential well describing the 
flake assumes a ``saw-tooth'' profile (see e.g. Ref. \cite{riss+11jacs}), 
whose height and orientation depends on the magnitude and on the sign of $\mu_x$. 
Hence the HOMO and LUMO states present a similar asymmetric character in both 
cases, but with spatially opposite localization (see \ref{fig3}b).
On the other hand, for the A-B scheme the $\pi$ distribution of the frontier 
orbitals undergoes only local modifications, specifically related to the altered 
symmetry of the flake\cite{symmetry-note}, as shown in \ref{fig3}c. 
The modified symmetry of the frontier orbitals is particularly evident for the 
\ce{NH2} termination, while it is less apparent for groups bonded through a C 
bond, such as \ce{COCH3}. See Supporting Information for a more complete analysis 
of the substituents and family dependence (Figure S1-S2).

\begin{table}
\begin{tabular}{c|c|c|c|c}
\hline \hline
Group & Scheme & $\mu_x$ [D] & $\mu_y$ [D] & $\mu_z$ [D] \\ \hline
 \multirow{2}{*}{\ce{F}} & A-A & 4.848 & 0.479 & 0.028 \\ \cline{2-5}
 & A-B & -0.001 & -0.003  & -0.061 \\ \hline 
 \multirow{2}{*}{\ce{NH2}} & A-A & -9.335 & -0.026 & -0.121 \\ \cline{2-5}
 & A-B & 0.339 & -1.238  & 3.823 \\ \hline 
 \multirow{2}{*}{\ce{COCH3}} & A-A & 9.764 & 1.138 & -0.049 \\ \cline{2-5}
 & A-B & 0.166 & -0.139  & 0.269 \\ \hline \hline
\end{tabular} 
\caption{
Components of the dipole moment for edge-functionalized $N$=7 GNFs (one substituent 
per class), according to the A-A and A-B functionalization schemes.
}
\label{table1}
\end{table}
%

%

%
\begin{table*}[t!]
\begin{tabular*}{0.98\textwidth}{l|c||cc|c|c||cc|c|c||cc|c|c}
\hline \hline
& \begin{scriptsize}\multirow{3}{*}{\textbf{H}}\end{scriptsize} & \multicolumn{4}{c||}{\begin{scriptsize}\textbf{Halogens}\end{scriptsize}} & \multicolumn{4}{c||}{\begin{scriptsize}\textbf{Direct bond}\end{scriptsize}} &  \multicolumn{4}{c}{\begin{scriptsize}\textbf{C-bond}\end{scriptsize}} \\ \cline{3-14}
 & & \multicolumn{2}{c|}{\begin{scriptsize}F\end{scriptsize}} & \begin{scriptsize}\multirow{2}{*}{Cl}\end{scriptsize} &  \begin{scriptsize}\multirow{2}{*}{Br}\end{scriptsize} & \multicolumn{2}{c|}{\begin{scriptsize}\ce{NH2}\end{scriptsize}} &  \begin{scriptsize}\multirow{2}{*}{\ce{OCH3}}\end{scriptsize}&  \begin{scriptsize}\multirow{2}{*}{\ce{OCF=CF2}}\end{scriptsize} &  
\multicolumn{2}{c|}{\begin{scriptsize}\ce{COCH3}\end{scriptsize}} &  \begin{scriptsize}\multirow{2}{*}{\ce{CH=CF2}}\end{scriptsize} &  \begin{scriptsize}\multirow{2}{*}{\ce{CH=CH2}}\end{scriptsize} \\ 
& & \begin{scriptsize}A-A\end{scriptsize} &  \begin{scriptsize}A-B\end{scriptsize} & & & \begin{scriptsize}A-A\end{scriptsize} &  \begin{scriptsize}A-B\end{scriptsize} & & & \begin{scriptsize}A-A\end{scriptsize} &  \begin{scriptsize}A-B\end{scriptsize} & & \\ \hline \hline
 \begin{scriptsize}\multirow{2}{*}{$\Delta$\textbf{EA}}\end{scriptsize} & \begin{scriptsize}\multirow{2}{*}{0.00}\end{scriptsize} & \begin{scriptsize}0.57\end{scriptsize} & \begin{scriptsize}0.58\end{scriptsize} & \begin{scriptsize}\multirow{2}{*}{0.53}\end{scriptsize} & \begin{scriptsize}\multirow{2}{*}{0.62}\end{scriptsize} &\begin{scriptsize} -0.18\end{scriptsize} & \begin{scriptsize}-0.14\end{scriptsize} &  \begin{scriptsize}\multirow{2}{*}{-0.12}\end{scriptsize} & \begin{scriptsize}\multirow{2}{*}{0.70}\end{scriptsize} & \begin{scriptsize}0.74\end{scriptsize} &\begin{scriptsize}0.69\end{scriptsize} & \begin{scriptsize}\multirow{2}{*}{0.67}\end{scriptsize} & \begin{scriptsize}\multirow{2}{*}{0.14}\end{scriptsize} \\ 
  & & \begin{scriptsize}\textit{0.01}\end{scriptsize} &\begin{scriptsize}\textit{0.01}\end{scriptsize} & & & \begin{scriptsize}\textit{0.08}\end{scriptsize} & \begin{scriptsize}\textit{0.04} \end{scriptsize}& & & \begin{scriptsize}\textit{0.01}\end{scriptsize} & \begin{scriptsize}\textit{0.01}\end{scriptsize} & &  \\ \hline
 \begin{scriptsize}\multirow{2}{*}{$\Delta$\textbf{IP}}\end{scriptsize} & \begin{scriptsize}\multirow{2}{*}{0.00}\end{scriptsize} & \begin{scriptsize}0.37\end{scriptsize} &\begin{scriptsize} 0.32\end{scriptsize} & \begin{scriptsize}\multirow{2}{*}{0.30} \end{scriptsize} &\begin{scriptsize} \multirow{2}{*}{0.39}\end{scriptsize} & \begin{scriptsize}-0.75\end{scriptsize} & \begin{scriptsize}-0.79\end{scriptsize} &  \begin{scriptsize}\multirow{2}{*}{-0.50} \end{scriptsize}& \begin{scriptsize}\multirow{2}{*}{0.31}\end{scriptsize} & \begin{scriptsize}0.49\end{scriptsize} & \begin{scriptsize}0.42\end{scriptsize} & \begin{scriptsize}\multirow{2}{*}{0.19}\end{scriptsize} & \begin{scriptsize}\multirow{2}{*}{-0.22}\end{scriptsize} \\ 
  & &\begin{scriptsize} \textit{-0.10}\end{scriptsize} & \begin{scriptsize}\textit{-0.10}\end{scriptsize} & & &\begin{scriptsize}\textit{-0.24}\end{scriptsize} & \begin{scriptsize}\textit{-0.19}\end{scriptsize} & & & \begin{scriptsize}\textit{-0.10}\end{scriptsize} & \begin{scriptsize}\textit{-0.10}\end{scriptsize} & &  \\ \hline
 \begin{scriptsize}\multirow{2}{*}{$E_G$}\end{scriptsize} & \begin{scriptsize}\multirow{2}{*}{5.13}\end{scriptsize} & \begin{scriptsize}4.92\end{scriptsize} & \begin{scriptsize}4.87\end{scriptsize} &\begin{scriptsize} \multirow{2}{*}{4.90}\end{scriptsize} & \begin{scriptsize}\multirow{2}{*}{4.90}\end{scriptsize} &\begin{scriptsize}4.55\end{scriptsize} & \begin{scriptsize}4.47 \end{scriptsize}&  \begin{scriptsize}\multirow{2}{*}{4.75}\end{scriptsize} & \begin{scriptsize}\multirow{2}{*}{4.74}\end{scriptsize} & \begin{scriptsize}4.88\end{scriptsize} & \begin{scriptsize}4.86\end{scriptsize} & \begin{scriptsize}\multirow{2}{*}{4.65}\end{scriptsize} & \begin{scriptsize}\multirow{2}{*}{4.76}\end{scriptsize} \\ 
  & & \begin{scriptsize}\textit{5.01} \end{scriptsize}& \begin{scriptsize}\textit{5.01}\end{scriptsize}& & & \begin{scriptsize}\textit{4.81}\end{scriptsize} & \begin{scriptsize}\textit{4.90}\end{scriptsize} & & & \begin{scriptsize}\textit{5.01}\end{scriptsize} & \begin{scriptsize}\textit{5.01}\end{scriptsize} & &  \\ \hline
\end{tabular*}
\caption{Differences of EA and IP ($\Delta$EA and $\Delta$IP, in eV) with respect 
to the reference hydrogenated (H) flake and energy gap ($E_G$, in eV) computed 
through AM1 model for $N$=7 GNFs, functionalized according to the A-B scheme. 
For selected groups, we report the values of $\Delta$EA, $\Delta$IP and $E_G$ 
also for the A-A scheme and for H-terminated model flakes with the corresponding 
distorted geometries (in italics).}
\label{table3}
\end{table*}

In \ref{table3} we report the differences of IP and EA ($\Delta$EA and $\Delta$IP)
of $N$=7 functionalized GNFs with respect to the reference hydrogenated flake,
as well as their energy gap ($E_G$).
As expected, an uneven shift of IP and EA is produced upon functionalization, related 
to the electronegativity of the substituents; for instance,  highly electronegative 
halogen terminations upwards shift EA more significantly than IP, while strongly 
electron-donating amino groups downwards move IP more than EA.
In any case, a gap reduction is always encountered. 
In the case of C-bonded terminations, a further contribution to the gap reduction
is given by the increased effective width of the flake, driven by the extended 
$\pi$-conjugation.

In addition to this well-established behavior, we here more directly estimate 
the role played by structural distortions. To quantify this effect, we consider 
model flakes obtained by removing the functionalizing groups from the final 
structure, and by then H-passivating the remaining dangling bonds: this model 
structure allows us to decouple the influence of distortions from the overall
influence of edge functionalization. 
By inspecting \ref{table3}, we find a total gap reduction upon functionalization
ranging from 200 to 700 meV. The comparison with the values of $E_G$ obtained 
for the model distorted H-GNFs (in italics in \ref{table3}) points to a non-%
negligible contribution of structural distortions, corresponding to 35-55$\%$ 
of the total amount. 
These results are very similar to what expected for thermal deviations from the 
ideal structure \cite{samo+10prb}.
It is also worth noting that this trend is mostly independent of the functionalization 
scheme, with a maximum difference of 80 meV for the \ce{NH2} case, where the two
schemes induce significantly different distortion patterns.

\begin{figure}
\centering
\includegraphics[width=.48\textwidth]{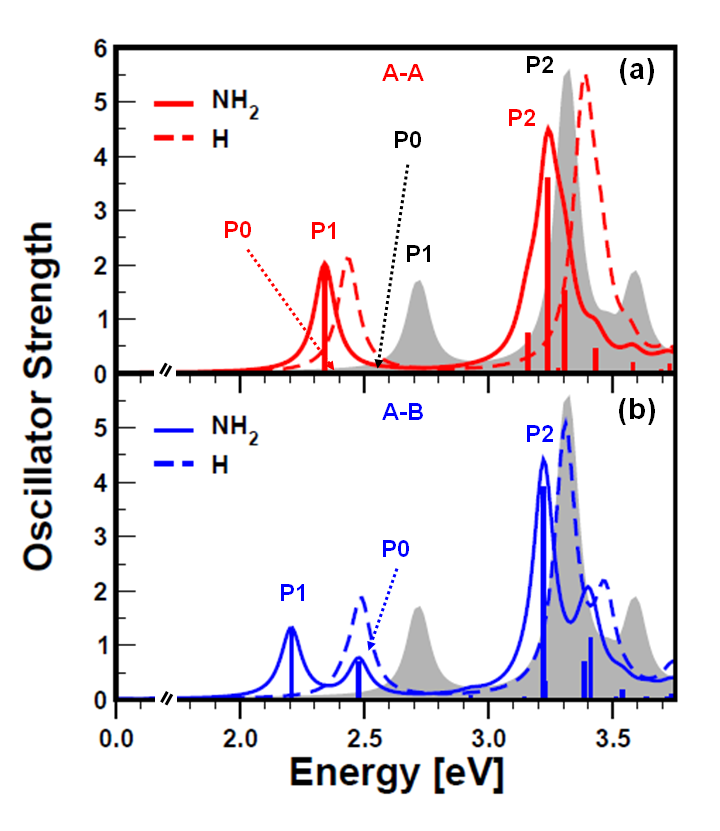}%
\caption{Optical spectra of GNFs of width parameter $N$=7, functionalized with 
\ce{NH2} groups, according to the (a) A-A and (b) A-B functionalization schemes.
In addition to the spectrum of the planar H-terminated flake taken as reference 
(grey shaded area, with the main excitations highlighted in a), we show the 
spectrum of model GNFs obtained by removing the edge substituents from the final 
structures and H-passivating the dangling bonds (dashed lines), in order to 
perform a rigorous analysis of distortions effects.
All the curves are obtained through a Lorentzian broadening of 50 meV.}%
\label{fig4}
\end{figure}

Starting from the AM1 results described above, we now turn to the analysis of
optical excitations, as obtained from ZINDO calculations.
To address the role of both molecular-symmetry and structural modifications, we 
first consider the $N$=7 GNF functionalized with \ce{NH2}, where these effects 
are both prominent.
In \ref{fig4} we report the optical spectra for both functionalization schemes, 
A-A (red) and A-B (blue), together with the spectrum of the flat H-GNF (grey 
shaded area), taken as a reference. In addition, we include the spectrum  
of the H-terminated distorted model flake described above (dashed line), which 
enables a direct analysis of the distortion effects. 
As already discussed \cite{cocc+12jpcl}, the optical spectrum of the planar $N$=7 
H-GNF shows two main peaks in the low-energy region with longitudinal polarization 
(along the $x$ axis), labelled as P1 and P2; in addition, a transversally polarized 
weak excitation (P0) is found at lower energy. 
Our results indicate that edge functionalization is responsible for a red shift 
of the P1 peak by 380 and 510 meV in the A-A (\ref{fig4}a) and A-B schemes 
(\ref{fig4}b), respectively.
Remarkably, a further peak is observed in the optical spectrum of the A-B 
functionalized flake, related to the oscillator strength (OS) increase of P0 by 
about 3 orders of magnitude.
This OS gain is a purely symmetry effect: the molecular symmetry reduction induced 
by functionalization (from D$_{2h}$ in the H-GNF to C$_{2h}$ in the A-B scheme) 
implies that the HOMO state acquires the same symmetry as the HOMO-1 one, so that 
the HOMO $\rightarrow$ LUMO transition is now allowed to contribute to the P0 
excitation (see \ref{table4}). 
This symmetry effect is in fact not observed in the A-A functionalized flake. 

Concerning the role of structural distortion, we observe that the largely distorted 
geometry produced by the A-A functionalizion (dashed line in \ref{fig4}a) is 
responsible for about 75$\%$ of the optical gap reduction, as noticed by comparing 
the excitation energy of P1 for the three curves in \ref{fig4}a.
P1 is now energetically degenerate with the dark excitation P0, also red shifted 
of about 200 meV with respect to the reference H-GNF (see \ref{table4}).
It is worth noting that the frontier orbital localization  observed for the A-A 
scheme in presence of polar functional groups (see \ref{fig3}b) does not affect 
the basic features of the main excitations: despite their spatial localization, 
the overlap between the frontier orbitals is still large enough to preserve the 
intensity of P1, in comparison with both the reference H-terminated and the A-B 
functionalized GNF.
In the case of the A-B functionalization pattern, distortions contribute to about 
45$\%$ of the whole optical gap red shift: the impact of distortions is less 
pronounced for this structure, and molecular symmetry effects dominate (see 
\ref{fig4}b).

\begin{table*}
\begin{tabular}{c|c|c|c|l}
\hline \hline \textbf{Termination}  & \textbf{Excitation}  & \textbf{Energy [eV]}  & \textbf{OS}  & \textbf{Transitions (weight)}  \\ 
\hline \hline  \multirow{5}{*}{H} & \multirow{2}{*}{P0} & \multirow{2}{*}{2.54} & \multirow{2}{*}{0.0005} &  H-1 $\rightarrow$ L (0.37) \\
& & & & H $\rightarrow$ L+1 (0.39) \\ \cline{2-5}
& P1 & 2.72 & 1.62 & H $\rightarrow$ L (0.79) \\ \cline{2-5}
& \multirow{2}{*}{P2} & \multirow{2}{*}{3.32} & \multirow{2}{*}{5.43} & H-3 $\rightarrow$ L+3 (0.12) \\
& & & & H-1 $\rightarrow$ L+1 (0.76) \\ 
\hline \multirow{6}{*}{F}  & \multirow{3}{*}{P0} & \multirow{3}{*}{2.43} & \multirow{3}{*}{0.21} &  H-1 $\rightarrow$ L (0.27) \\
& & & & H $\rightarrow$ L (0.12) \\
& & & & H $\rightarrow$ L+1 (0.37) \\ \cline{2-5}
& P1 & 2.58 & 1.54 & H $\rightarrow$ L (0.68) \\ \cline{2-5}
&  \multirow{2}{*}{P2} & \multirow{2}{*}{3.26} & \multirow{2}{*}{5.11} &  H-3 $\rightarrow$ L+3 (0.20) \\
& & & & H-1 $\rightarrow$ L+1 (0.71) \\
\hline \hline \multirow{4}{*}{\ce{NH2} -- A-A} & P1 & 2.34 & 1.98 &  H $\rightarrow$ L (0.75) \\ \cline{2-5}
& \multirow{2}{*}{P0} & \multirow{2}{*}{2.35} & \multirow{2}{*}{0.02} &  H-2 $\rightarrow$ L (0.29) \\
& & & & H $\rightarrow$ L+1 (0.39) \\ \cline{2-5}
& P2 & 3.24 & 3.59 &  H-2 $\rightarrow$ L+1 (0.50) \\
\hline \multirow{6}{*}{\ce{NH2} -- A-B} & P1 & 2.21 & 1.29 &  H $\rightarrow$ L (0.63) \\ \cline{2-5}
& \multirow{3}{*}{P0} & \multirow{3}{*}{2.48} & \multirow{3}{*}{0.70} &  H-2 $\rightarrow$ L (0.22) \\
& & & & H $\rightarrow$ L (0.18) \\
& & & & H $\rightarrow$ L+1 (0.40) \\ \cline{2-5}
&  \multirow{2}{*}{P2} & \multirow{2}{*}{3.22} & \multirow{2}{*}{3.92} &  H-2 $\rightarrow$ L+1 (0.53) \\
& & & & H $\rightarrow$ L+1 (0.10) \\
\hline \hline \multirow{4}{*}{\ce{COCH3}} & \multirow{2}{*}{P0} & \multirow{2}{*}{2.42} & \multirow{2}{*}{0.04} &  H-1 $\rightarrow$ L (0.34) \\
& & & & H $\rightarrow$ L+1 (0.40) \\ \cline{2-5}
& P1 & 2.55 & 1.68 & H $\rightarrow$ L (0.78) \\ \cline{2-5}
& P2 & 3.21 & 5.13 & H-1 $\rightarrow$ L+1 (0.71) \\
\hline \hline
\end{tabular} 
\caption{Energy, oscillator strength (OS) and composition in terms of molecular 
orbital transitions of the main excitations of functionalized $N$=7 GNFs, in 
presence of one prototypical substituent per class, i.e. F, \ce{NH2} (also A-A 
scheme) and \ce{COCH3}, in addition to the reference hydrogenated system. 
In the last column we report the MO transitions contributing to each excitation
which have a relative weight larger than 0.1}
\label{table4}
\end{table*}
%


\begin{figure}
\centering
\includegraphics[width=.48\textwidth]{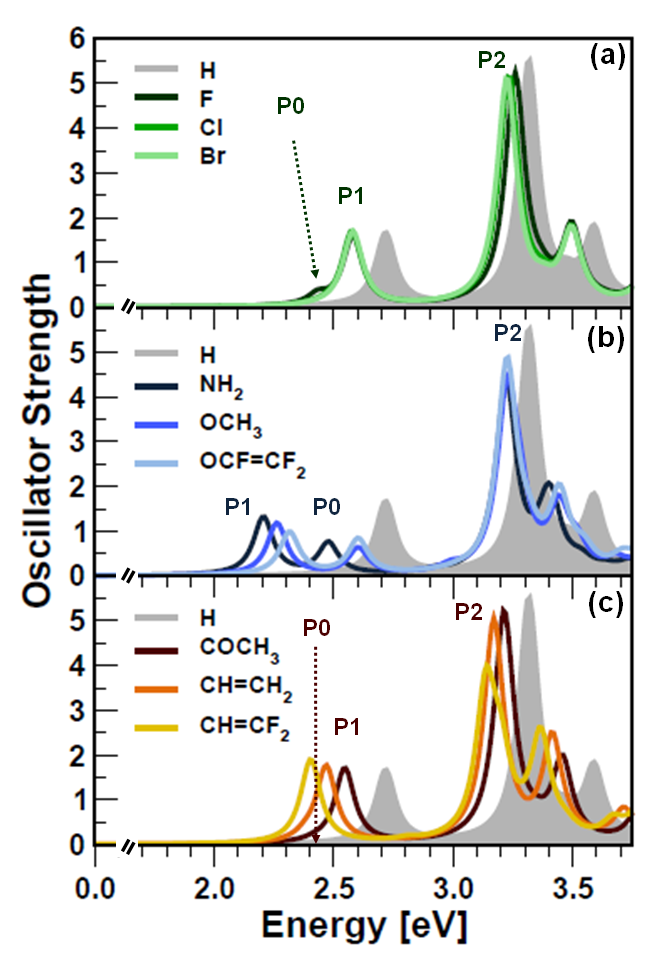}%
\caption{Optical spectra N=7 GNFs functionalized in the A-B scheme
with each terminations considered in this work, including (a) halogen atoms, (b) 
directly- and (c) C-bonded groups. 
In each panel, we indicate for one of the groups the three main excitations 
described in the text, i.e. P0, P1 and P2. Moreover, the spectrum of the reference 
H-flake is reported (grey shaded area in each panel). 
All the curves are obtained through a Lorentzian broadening of 50 meV.}%
\label{fig5}
\end{figure}

We next inspect the role of the chemical specificities of edge substituents, by 
comparing the optical spectra of $N$=7 GNFs functionalized with all the anchoring 
groups considered in this work, as shown in \ref{fig5}.
We here consider only the A-B pattern, since negligible differences are observed 
between the two schemes for most of the substituents (see Figure S3 in the 
Supporting Information for more details). The analysis of fully halogen terminated 
GNFs is also included in the Supporting Information (see Figure S4 and Table S8).
A non rigid red shift of the spectra is generally observed upon functionalization, 
with a larger red shift for P1 than for P2.  
All the spectra of the groups anchored through a direct bond (\ref{fig5}b) display 
the same key feature already discussed for the amino termination and reported in 
\ref{fig4}b: an additional distinct peak emerges, corresponding to the P0 
excitation which becomes optically active.
This effect is also present, even though less intense, for halogen terminations:
P0 appears as a shoulder in the main P1 peak for fluorinated flakes (\ref{fig5}a).
In the case of C-bonded molecular substituents, the impact of functionalization  
on the electronic states is further reduced, so that the main spectral features  
are basically preserved with respect to the flat H-terminated flake (see spectra 
in \ref{fig5}b and the composition of the main excitations in \ref{table4}).

\begin{figure*}
\centering
\includegraphics[width=.9\textwidth]{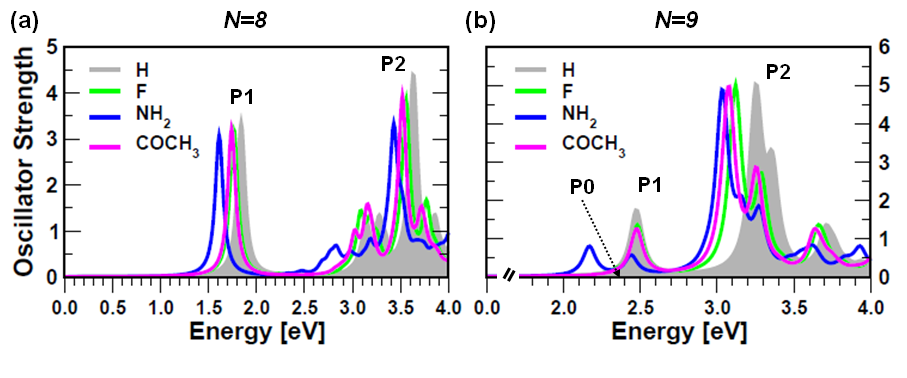}%
\caption{Optical spectra of symmetrically functionalized GNFs of width parameter 
(a) $N$=8 (A-A scheme) and (b) $N$=9 (A-B scheme). One edge substituent for each class
 of terminations considered in this work is addressed for both GNFs. The spectrum of the reference
 H-GNF is reported in the background of each panel (grey shaded area) with the indication of the main 
excitations (P0, P1 and P2). All the curves are obtained through a Lorentzian broadening of 50 meV.}%
\label{fig6}
\end{figure*}

In order to complete our analysis, we finally investigate the spectra of the GNFs 
of width parameter $N=8$ and $N=9$, functionalized with one prototypical group 
per class, namely F, \ce{NH2} and \ce{COCH3}, according to the higher symmetry 
scheme (A-A for $N$=8 and A-B for $N$=9).
The optical spectra computed for these systems are shown in \ref{fig6}, with the 
reference spectrum of the corresponding planar H-GNF reported in the background 
of each panel (grey shaded area).
The optical spectrum of the $N$=8 GNF is dominated by a first intense peak P1, 
which is about 1 eV lower in energy compared to those of the other GNF families 
(compare \ref{fig6}a with \ref{fig5} and \ref{fig6}b), in agreement with the 
trend for the electronic gaps \cite{cocc+11jpcc}. 
Contrary to $N=$7, this is always the lowest energy excitation, regardless the 
edge termination (See Supporting Information, Table S5).
Moreover, we do not observe any functionalization-induced modification in the 
$\pi$ distribution of frontier orbitals, since the molecular symmetry remains the 
same in the A-A scheme (C$_{2h}$). In this case the main effect of functionalization 
is thus to red shift the spectrum with respect to the reference H-GNF, in an almost 
rigid fashion (up to over 200 meV upon amino termination).

On the other hand, the optical spectra of $N$=9 GNFs manifest more similarities 
with those of $N$=7 GNFs (see \ref{fig6}b and \ref{fig5} for comparison), both in 
the spectral shape and in the composition of the main excitations (see Supporting 
Information, Table S6). Also for the ribbons belonging 
to the $N$=3$p$ family, the P0 excitations becomes optically active, giving rise 
to a distinct peak in presence of \ce{NH2} functional groups, according to the 
mechanism described for the $N$=7 GNFs.
The lowest energy peak is red shifted by about 300 meV upon amino substitution, 
compared to the reference H-GNF.
In presence of other edge substituents (F and \ce{COCH3}), the energy of P1 is 
basically unaffected, while P2 is red shifted of about 200 meV upon each termination 
(see \ref{fig6}b).
This is an effect of the MO distribution observed for the GNFs belonging to this 
family, for which the frontier orbitals present reversed orientation of the $\pi$ 
distribution with respect to the flakes of the $N$=3$p$+1 family
(see Figure S2).

\section{Conclusions}
We have analyzed in details the effects of edge covalent functionalization on the
structural, electronic and optical properties of armchair graphene nanoflakes. 
Different functionalization groups and patterns have been discussed for a specific 
flake, and the analysis extended and rationalized for additional systems, according to 
the armchair width modulation.
Our results indicate that steric repulsion among substituents generally determines 
local distortion at the edges, even though structural modifications can involve the 
whole flake backbone in specific cases.
Functionalization-induced distortions are found to play an effective role in 
the electronic gap reduction experienced by the flakes, with contributions 
ranging from 35 to 55$\%$ of the total gap decrease. 
Depending on the functionalization pattern and the character of the substituents, 
interesting charge-redistribution effects are seen for the frontier orbitals, 
either related to the modified molecular symmetry or to the appearance of a 
large longitudinal dipole component.
All these effects are reflected in the optical spectra: distortions largely 
contribute to the lowering of the optical gap (up to 300 meV), while the appearance
of an additional peak at low energy is related to the distribution and the 
characteristics of the functional groups. 
Our results indicate that the interplay among the system symmetry, the chemical 
features and the distribution of the substituents, as well as the resulting 
structural distortions is crucial to understand the modification of the opto-%
electronic properties of functionalized graphene nanoflakes.
\begin{acknowledgement}
The authors are grateful to Stefano Corni for fruitful discussion and acknowledge 
CINECA for computational support.
This work was partly supported by the Italian Ministry of University and Research 
under FIRB grant ItalNanoNet, and by Fondazione Cassa di Risparmio di Modena with 
project COLDandFEW.
M.~J.~C.~ acknowledges support from FAPESP and CNPq (Brazil).
\end{acknowledgement}
\begin{suppinfo}
The Supporting Information is organized in three sections.
In the first one we report the structural details (bond lengths and distortion 
angles) of the optimized $N$=7, $N$=8 and $N$=9 graphene nanoflakes, to 
complete the data reported in the main text.
The second section supports the analysis on the electronic properties, including 
a table with the dipole moment components of $N$=8 and $N$=9 GNFs (functionalized 
according to both A-A and A-B schemes) and the isosurfaces of the frontier orbitals
for the considered systems.
The last part is dedicated to optical properties.
The analysis on the effects related to different functionalization schemes and 
structural distortions are extended here also to \ce{COCH3} and F terminations, 
as representatives of the classes of C-bonded and halogen substituents.
We also include the tables with the composition of the main excitations for 
$N$=8 and $N$=9 GNFs, and the analysis of the optical features of full halogen%
-terminated GNFs (only $N$=7).
\end{suppinfo}
\providecommand*{\mcitethebibliography}{\thebibliography}
\csname @ifundefined\endcsname{endmcitethebibliography}
{\let\endmcitethebibliography\endthebibliography}{}

\clearpage


\section*{Supplementary material}

In Table S5 and Table S6 we report the composition of the main excitations of functionalized GNFs of width parameters $N=8$ and $N=9$, respectively. The corresponding spectra are shown in the main text (Figure 6).

**Table S5: Energy, oscillator strength (OS) and composition of the two lowest energy excitations of functionalized GNFs of width parameter $N=8$ (A-A scheme, see spectra in Figure 6a of the main text). In the last column only the molecular orbital transitions with relative weights larger than 0.1 are reported.**

| System | Excitation | Energy [eV] | OS | Transitions (weight) |
|---|---|---|---|---|
| H | P1 | 1.84 | 3.33 | H $\to$ L (0.91) |
| | P2 | 3.64 | 3.47 | H-2 $\to$ L+2 (0.56) |
| F | P1 | 1.76 | 3.22 | H $\to$ L (0.91) |
| | P2 | 3.56 | 3.62 | H-2 $\to$ L+2 (0.62) |
| NH$_2$ | P1 | 1.69 | 3.04 | H $\to$ L (0.91) |
| | P2 | 3.04 | 1.64 | H-2 $\to$ L (0.46)<br>H $\to$ L+2 (0.18)<br>H $\to$ L+4 (0.15) |
| COCH$_3$ | P1 | 1.73 | 3.22 | H $\to$ L (0.91) |
| | P2 | 3.14 | 1.08 | H-2 $\to$ L+1 (0.23)<br>H-2 $\to$ L (0.10)<br>H $\to$ L+4 (0.26)<br>H $\to$ L+6 (0.13) |

Table S6: **Energy, oscillator strength (OS) and composition of the two lowest energy excitations of functionalized GNFs of width parameter** $N$=9 **(A-B scheme, see spectra in Figure 6b of the main text). In the last column only the molecular orbital transitions with relative weights larger than 0.1 are reported.**

| System | Excitation | Energy [eV] | OS | Transitions (weight) |
|---|---|---|---|---|
| H | P0 | 2.40 | 0.00 | H-1 → L (0.37)<br>H → L+1 (0.41) |
|  | P1 | 2.47 | 1.71 | H → L (0.85) |
|  | P2 | 3.24 | 4.47 | H-2 → L+2 (0.12)<br>H-1 → L+1 (0.73) |
| F | P0 | 2.34 | 0.06 | H-1 → L (0.36)<br>H → L+1 (0.39) |
|  | P1 | 2.48 | 1.31 | H → L (0.78) |
|  | P2 | 3.12 | 4.80 | H-1 → L+1 (0.69) |
| NH$_2$ | P1 | 2.17 | 0.76 | H → L (0.55)<br>H → L+1 (0.12) |
|  | P0 | 2.44 | 0.50 | H-1 → L (0.27)<br>H → L (0.24)<br>H → L+1 (0.33) |
|  | P2 | 3.03 | 4.64 | H-1 → L+1 (0.57) |
| COCH$_3$ | P0 | 2.33 | 0.01 | H-1 → L (0.37)<br>H → L+1 (0.41) |
|  | P1 | 2.48 | 1.21 | H → L (0.80) |
|  | P2 | 3.08 | 4.55 | H-1 → L+1 (0.64) |

We finally present the results on the optical properties of fully functionalized $N=7$ GNFs with halogen terminations. In Figure S4(a) we show the computed spectra, where the main excitations are indicated for the reference H-GNF (grey shaded area). In case of full functionalization the model symmetry of the H-terminated flake is preserved and hence the character of the frontier orbitals [see Figure S4(b)] as well as the composition of the main excitations in terms of molecular orbital transitions and their intensity are not significantly affected (see Table S7). A red shift of the spectra (about 350 meV for P1 and about 250 meV for P2) is noticed, compared to the reference H-GNF.

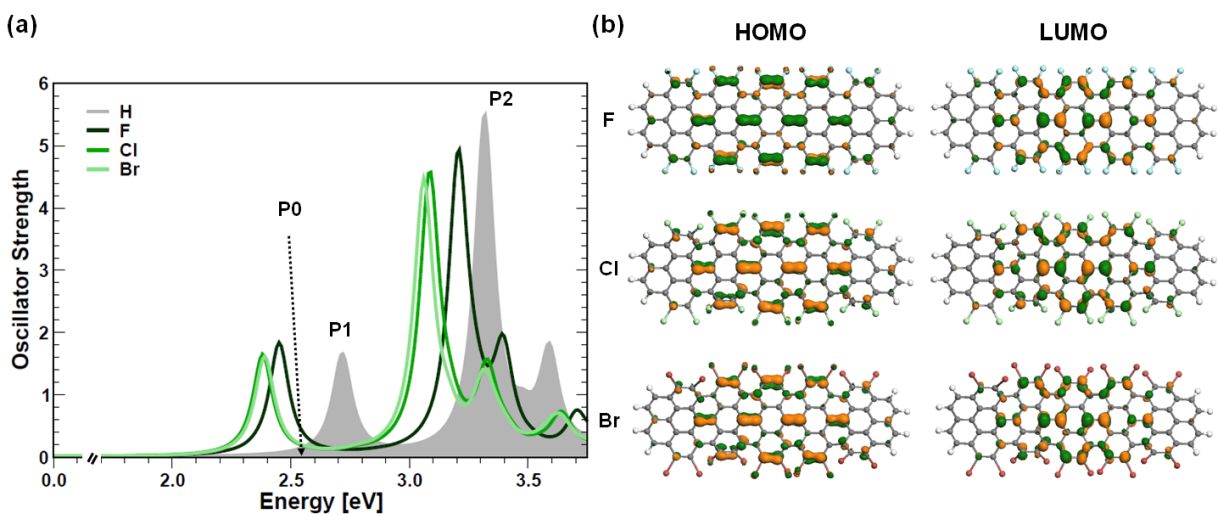

Figure S4: (a) Optical spectra of $N=7$ graphene nanoflakes fully functionalized with halogen atoms. For comparison, the spectrum of the reference H-terminated flake is included (grey shaded area with the labels of the main excitations). (b) Frontier orbitals of the corresponding structures.

Table S7: Energy, oscillator strength (OS) and composition of the main excitations (P0, P1 and P2) of fully functionalized $N$=7 GNFs with halogen terminations (F, Cl, Br – see Figure S5). In the last column only the molecular orbital transitions with relative weights larger than 0.1 are reported.

| System | Excitation | Energy [eV] | OS | Transitions (weight) |
|---|---|---|---|---|
| H | P0 | 2.54 | 0.0005 | H-1 → L (0.37)<br>H → L+1 (0.39) |
| H | P1 | 2.72 | 1.62 | H → L (0.79) |
| H | P2 | 3.32 | 5.43 | H-3 → L+3 (0.12)<br>H-1 → L+1 (0.76) |
| F | P0 | 2.36 | 0.02 | H-1 → L (0.30)<br>H → L+1 (0.45) |
| F | P1 | 2.45 | 1.81 | H → L (0.82) |
| F | P2 | 3.21 | 4.70 | H-3 → L+3 (0.11)<br>H-2 → L+1 (0.73) |
| Cl | P0 | 2.27 | 0.01 | H-1 → L (0.28)<br>H → L+1 (0.42) |
| Cl | P1 | 2.38 | 1.62 | H → L (0.81) |
| Cl | P2 | 3.08 | 4.42 | H-3 → L+2 (0.10)<br>H-1 → L+1 (0.67) |
| Br | P0 | 2.27 | 0.01 | H-1 → L (0.29)<br>H → L+1 (0.42) |
| Br | P1 | 2.39 | 1.59 | H → L (0.81) |
| Br | P2 | 3.06 | 4.33 | H-3 → L+2 (0.10)<br>H-1 → L+1 (0.69) |




\begin{mcitethebibliography}{62}
\bibitem[sym()]{symmetry-note}
 The HOMO (LUMO) level of the $N$=7 H-terminated flake (D$_{2h}$ point group)
  has B$_{3u}$ (B$_{2g}$) symmetry. Within the A-B functionalization scheme,
  the flake symmetry reduces to C$_{2h}$ and consequently the HOMO (LUMO)
  becomes a A$_u$ (B$_g$) state. It is worth recalling that the symmetry
  considerations reported here are for ideal unrelaxed systems. Structural
  optimization further lowers the symmetry of functionalized flakes and
  consequently locally affects also the molecular orbital distribution.\relax
\mciteBstWouldAddEndPunctfalse
\mciteSetBstMidEndSepPunct{\mcitedefaultmidpunct}
{}{\mcitedefaultseppunct}\relax
\EndOfBibitem
\end{mcitethebibliography}
\end{document}